\documentclass[twocolumn,showpacs,preprintnumbers,amsmath,amssymb,prl]{revtex4}

\usepackage{graphicx}
\usepackage{bm}

\begin{document}

\title{Quasi-binary amorphous phase in a 3D system of particles with  repulsive-shoulder interactions}

\author{Yu. D. Fomin}
\affiliation{Institute for High Pressure Physics, Russian Academy
of Sciences, Troitsk 142190, Moscow Region, Russia}

\author{Daan Frenkel}
\affiliation{FOM Institute for Atomic and Molecular Physics,
Amsterdam, The Netherlands and Dept. of Chemistry, Univ. of
Cambridge, Cambridge, UK}

\author{N. V. Gribova}
\affiliation{Institute for High Pressure Physics, Russian Academy
of Sciences, Troitsk 142190, Moscow Region, Russia}

\author{V. N. Ryzhov}
\affiliation{Institute for High Pressure Physics, Russian Academy
of Sciences, Troitsk 142190, Moscow Region, Russia}

\author{S. M. Stishov}
\affiliation{Institute for High Pressure Physics, Russian Academy
of Sciences, Troitsk 142190, Moscow Region, Russia}

\date{\today}

\begin{abstract}
We report a  computer-simulation study of the equilibrium phase
diagram of a three-dimensional system of particles with a
repulsive step potential. Using free-energy calculations, we have
determined the equilibrium phase diagram of this system. At low
temperatures, we observe a number of distinct crystal phases.
However, under certain conditions the system undergoes a glass
transition in a regime where the liquid appears thermodynamically
stable.  We argue that the appearance of this amorphous
low-temperature phase can be understood by viewing this
one-component system as a pseudo-binary mixture.
\end{abstract}

\pacs{61.20.Gy, 61.20.Ne, 64.60.Kw} \maketitle %

Many ordered materials can undergo a first-order phase transition
that does not alter the symmetry of the phase. In the early work
of Hemmer and Stell \cite{stell} it was proposed that such
iso-structural phase transitions are to be expected if the
interaction between the particles in addition to the ordinary hard
core has a ``soft'' core combined with an attractive interaction
at larger distances. The work of ref.~\cite{stell} focused
specifically on iso-structural phase transitions in materials such
as Ce or Cs, but since then many authors have studied a wide range
of model system that could exhibit iso-structural
transitions~\cite{book1,malescio,stell1,RS2002,FRT2006,stanley98,
stanley99, jagla1,young1,fren97,stishov}. Most of these
systems have a repulsive intermolecular potential that has a
region of negative curvature, a feature that is known to be
present in the interatomic potentials of some pure metallic
systems, metallic mixtures, electrolytes and colloidal systems.
The simplest example of a negative-curvature potential is the
repulsive-step potential which consists of a hard core plus a
finite repulsive shoulder at a larger radius. Systems of particles
interacting through such pair potentials  can possess a rich
variety of phase transitions and thermodynamic anomalies,
including liquid-liquid phase transitions \cite{RS2002,FRT2006},
water-like anomalies \cite{stanley98}, and isostructural
transitions in the solid region \cite{young1,fren97}.

The repulsive step potential has the form:
\begin{equation}
\Phi (r)=\left\{
\begin{array}{lll}
\infty , & r\leq d \\
\varepsilon , & d <r\leq \sigma  \\
0, & r>\sigma%
\end{array}%
\right.  \label{1}
\end{equation}
where $d$ is the diameter of the hard core, $\sigma$ is the width
of the repulsive step,  and  $\varepsilon$ its height. In the
low-temperature limit $\tilde{T}\equiv k_BT/\varepsilon<<1$  the
system reduces to a hard-sphere systems with hard-sphere diameter
$\sigma$, whilst in the limit $\tilde{T}>>1$ the system reduces to
a hard-sphere model with a hard-sphere diameter $d$. For this
reason, melting at high and low temperatures follows simply from
the hard-sphere melting curve $P=cT/\sigma'^3$, where $c \approx
12$ and $\sigma'$ is the relevant hard-sphere diameter ($\sigma$
and $d$, respectively). A changeover from the low-$T$ to high-$T$
melting behavior should occur for $\tilde{T} ={\mathcal O}(1)$.
The precise form of the phase diagram depends on the ratio
$s\equiv \sigma/d$. For large enough values of $s$ one should
expect to observe in the resulting melting curve a maximum that
should disappear as $s\rightarrow 1$ \cite{stishov}. The phase
behavior in the crossover region may be very complex, as  shown
below.

In our simulations we have used a smoothed version of the
repulsive step potential (Eqn.~\ref{1}), which has the form:
\begin{equation}
\Phi (r)=
\left(\frac{d}{r}\right)^{n}+\frac{1}{2}\varepsilon\left(1-\tanh\left(k_0
\left(r-\sigma_s \right)\right)\right) \label{2}
\end{equation}
where $n=14,k_0=10$. We have considered the following values of
$\sigma_s$: $\sigma_s=1.15,1.35$. In the remainder of this paper
we use the dimensionless quantities: $\tilde{{\bf r}}\equiv {\bf
r}/d$, $\tilde{P}\equiv P d^{3}/\varepsilon ,$ $\tilde{V}\equiv
V/N d^{3}\equiv 1/\tilde{\rho}$. As we will only use these reduced
variables, we omit the tildes.

\begin{figure}
\includegraphics[width=6cm]{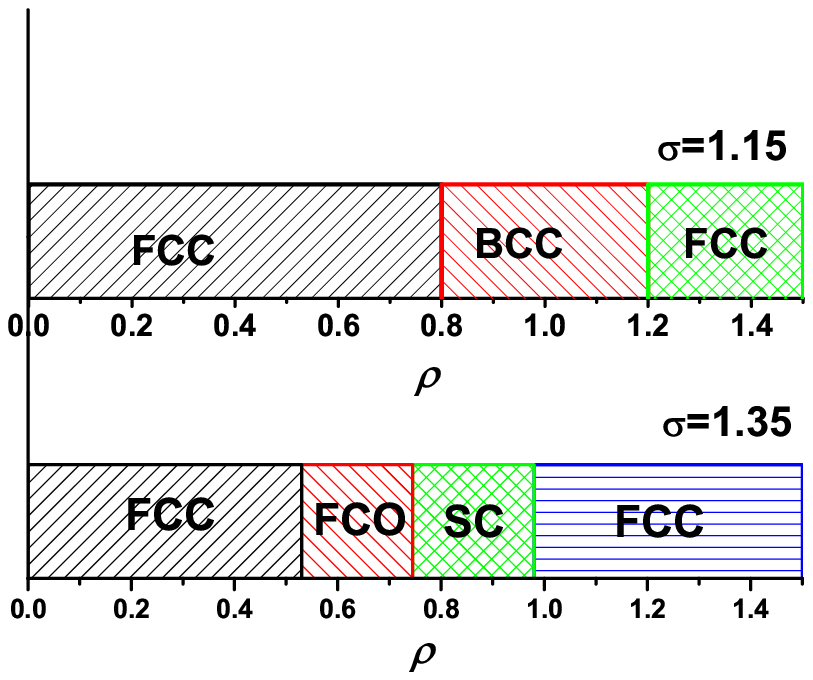}%
\caption{\label{fig:fig2} Ground state. The schematic phase
diagrams at $T=0$ for $\sigma_s=1.15$ and $\sigma_s=1.35$.}
\end{figure}

In order to get a hint about the phase diagrams of the system we
computed energies per particle for different crystal ground states
for $\sigma_s=1.15, 1.35$. In the upper part of
Fig.~\ref{fig:fig2} we plot a schematic phase diagram at $T=0$ for
$\sigma_s=1.15$. It is obtained from the calculation of the ground
state energy of the FCC (face-centered cubic) and BCC
(body-centered cubic) crystal structures.
 We also  considered  several other structures for this potential
 - SC (simple cubic), HCP, diamond structure, FCO (face-centered orthorhombic)
 and BCO  (body-centered orthorhombic) but all these were
 unstable.

In the lower part of Fig.~\ref{fig:fig2} we show the schematic
phase diagram at $T=0$ for $\sigma_s=1.15$. It is plotted from the
calculation of the ground state energies of the FCC, BCC, SC,
diamond and FCO structures for $\sigma_s=1.35$.
Fig.~\ref{fig:fig2} shows that there exists a range of densities
where the stable phase has an FCO structure whose unit cell is
increased along the $z$ axis in 1.6 times.
This is a clear indication that the phase diagrams in the solid
region are quite complex.

To determine the phase diagram at non-zero temperature, we
performed  constant-NVT MD simulations combined with free-energy
calculations. In all cases, periodic boundary conditions were
used. The number of particles varied between 250 and 500. No
system-size dependence of the results was observed. The system was
equilibrated for $5\times 10^6$ MD time steps. Data were
subsequently collected during $3\times 10^6\delta t$ where the
time step $\delta t=5\times 10^{-5}$.

In some region of the phase diagram for $\sigma_s=1.35$ we found
that the liquid phase was thermodynamically stable down to the
lowest temperatures that we could probe.   In this region of the
phase diagram, the dynamics of the system was very sluggish,
forcing us to perform much longer MD simulations (up to $90\times
10^6$ MD steps).

In order to map out the phase diagram of the system, we computed
its Helmholtz free energy using the thermodynamic integration: the
free energy of the liquid phase was computed via thermodynamic
integration from the dilute gas limit~\cite{book_fs}, and the free
energy of the solid phase was computed by thermodynamic
integration to an Einstein crystal~\cite{book_fs,fladd}.  In the
MC simulations of solid phases, data were collected during $5
\times 10^4$ cycles after equilibration. To improve the statistics
(and to check for internal consistency) the free energy of the
solid was computed at many dozens of different state-points. All
free-energy data were used to construct a single multinomial fit
for the equation of state in every phase. The transition points
were determined by a double-tangent construction (DTC).

\begin{figure}
\includegraphics[width=5.5cm]{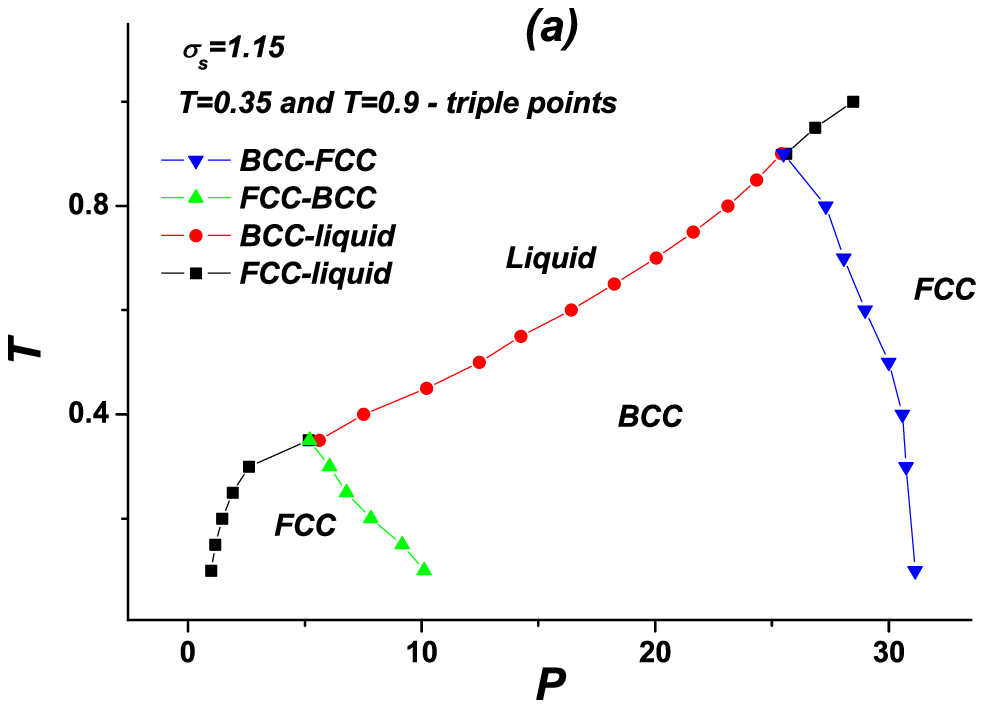}%

\includegraphics[width=6cm]{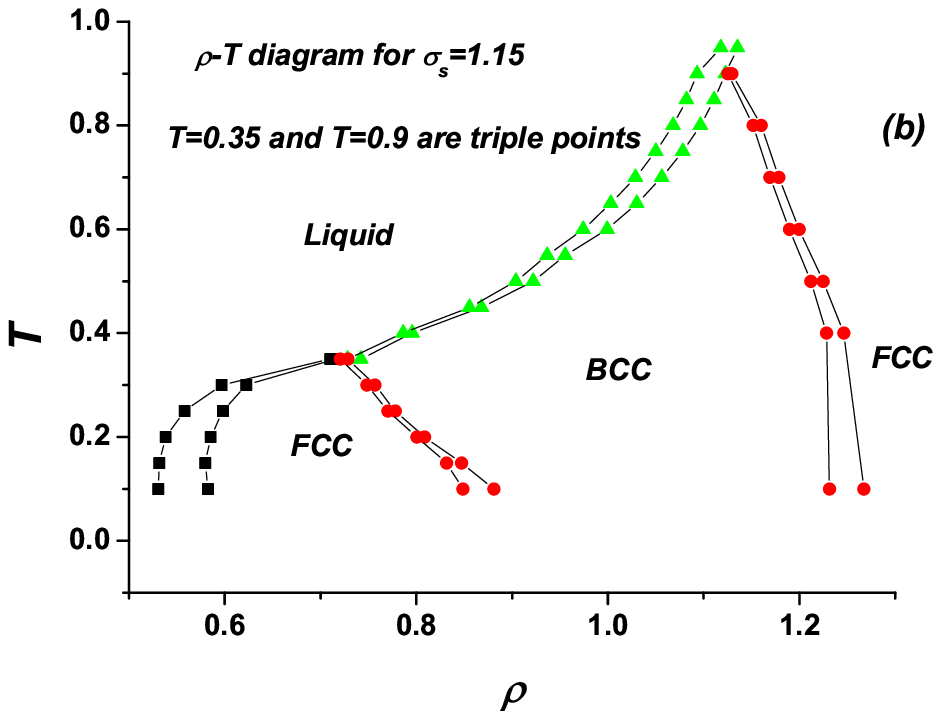}%

\includegraphics[width=6cm]{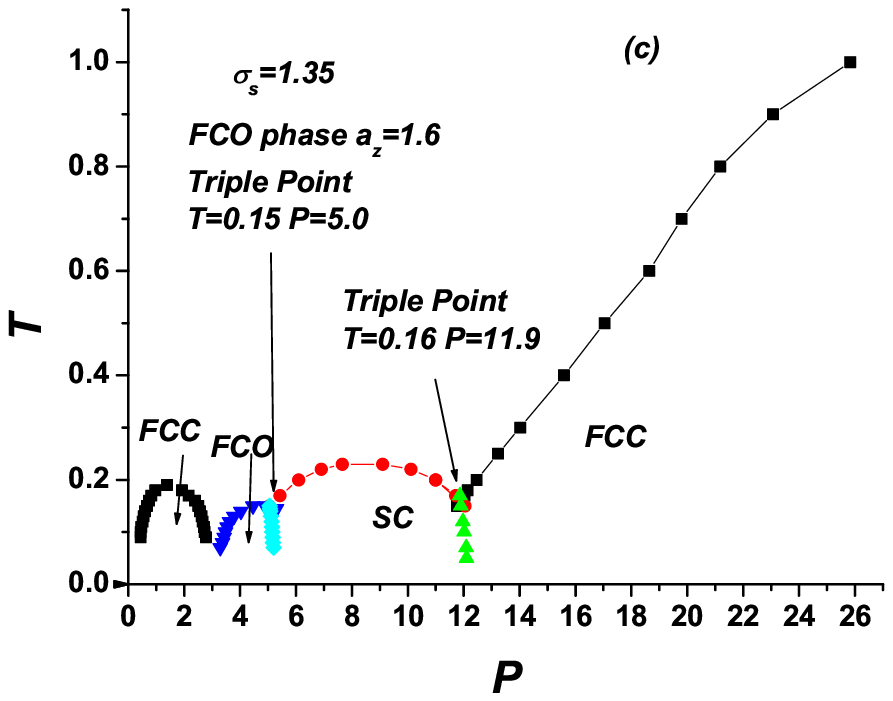}%

\includegraphics[width=6cm]{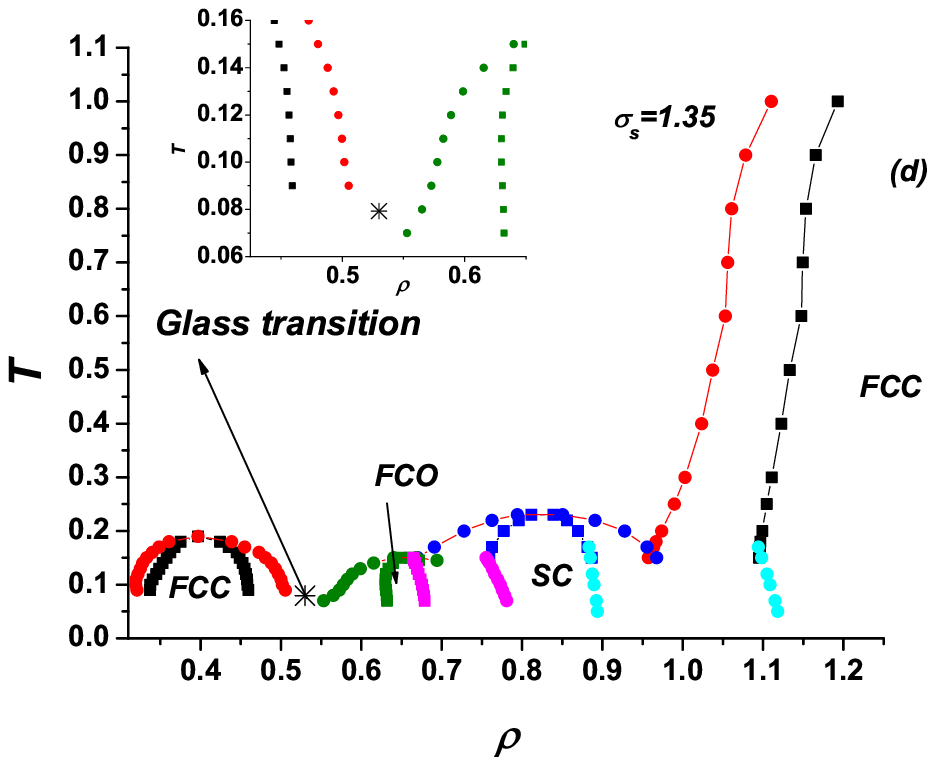}%
\caption{\label{fig:fig4} Phase diagram of the system of particles
interacting through the potential (2) with $\sigma_s=1.15, 1.35$
in $P-T$ and $\rho-T$ planes. The asterisk in Fig. (d) corresponds
to the glass transition.}
\end{figure}

Fig.~\ref{fig:fig4} shows the phase diagrams that we obtain from
the free-energy calculations for two different values of
$\sigma_s$. Figs.~\ref{fig:fig4}(a-b) show the phase diagram of
the system with $\sigma_s=1.15$.  One can see that for the system
with $\sigma_s=1.15$ there are no maxima in the melting curve. In
a soft-sphere system described by the potential $1/r^{14}$ a
face-centered cubic crystal structure has been
reported~\cite{Kofke}. However, the addition of a small repulsive
step leads to the appearance of the FCC-BCC transition shown in
Figs.~\ref{fig:fig4}(a-b).

Figs.~\ref{fig:fig4}(c-d) show the phase diagram of the system
with $\sigma_s=1.35$ in the $P-T$ and $\rho-T$ planes. There is a
clear maximum in the melting curve at low densities. The phase
diagram consists in two isostructural FCC parts corresponding to
close packing of the small and large spheres separated by a
sequence of structural phase transitions.

This complexity of the phase diagram can be understood from the
shape of the potential. Since the potential is purely repulsive,
the particles tend to minimize  both hard and soft-core overlaps.
Such overlaps can be avoided in the low pressure part of the phase
diagram. Upon increasing the density, the system crystallizes into
FCC phase which corresponds to the close packing of the particles
at the soft core.  If the pressure increases more, the particles
begin to penetrate the soft core but still avoid hard-core
overlaps.  Hence, both parts of the potential are now important.
The FCO phase corresponds to the situation when the particles are
packed in $X-Y$ planes in accordance with the hard core but the
planes are arranged at the soft core distance (layered structure).
As a result a particle has 4 nearest neighbors in this structure.
If we further increase the pressure, more particles penetrate the
soft core, but still the closed packed structures are not
favorable. Finally at very  high pressures region we observe the
FCC phase corresponding to the $r^{-14}$ limit.

\begin{figure}
\includegraphics[width=6cm]{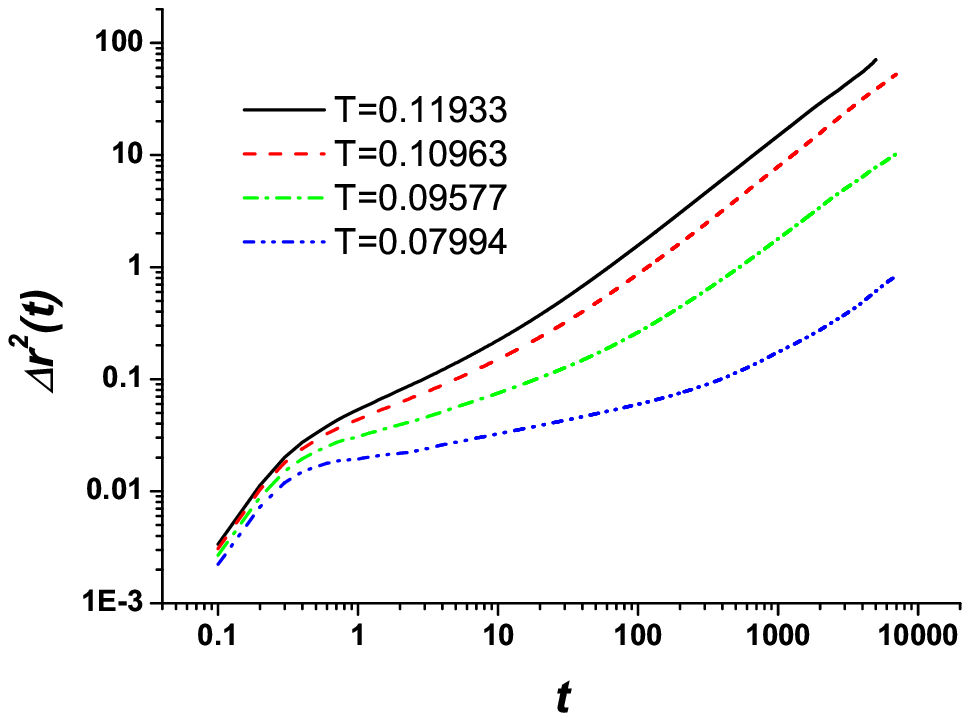}

\includegraphics[width=6cm]{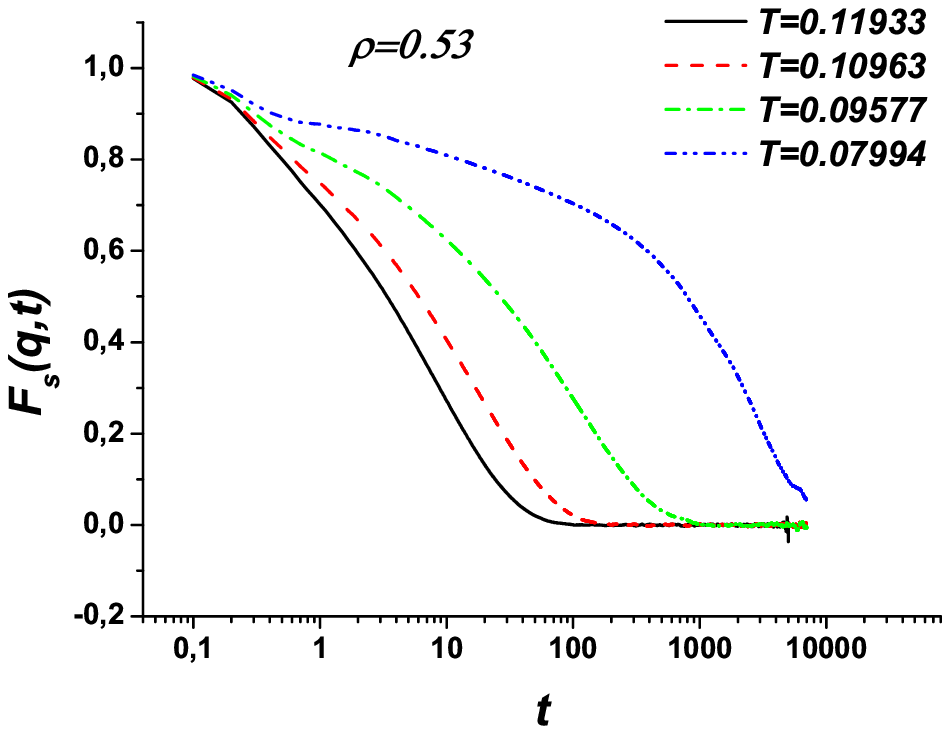}%
\caption{\label{fig:fig3} (a) MSD and (b) self-correlation
function $F_s(q,t)$ as functions of time for $\sigma=1.35$ for
different temperatures at $\rho=0.53$.}
\end{figure}

Interestingly, there is a region of the phase diagram where we
have not found any stable crystal phase.
 In order to clarify the properties
of the system in the "gap" on the phase diagram between the
low-density FCC phase and FCO phase, we compute mean-squared
displacement (MSD, $\Delta r^2(t)$) and self-correlation function
$F_s(q,t)$ for different temperatures at $\rho=0.53$
(Fig.~\ref{fig:fig3}). As usually observed in the proximity of
liquid-glass transition \cite{KobAndersen,Kob,Moreno}, a bending
occurs in the MSD after the initial ballistic regime
(Fig.~\ref{fig:fig3}(a)). A plateau appears at low temperatures
which corresponds to the onset of the caging regime. At long time,
the diffusion regime ($ \Delta r^2(t)\propto t$) is reached, when
the particles move, on average, a distance of the order of their
size.

Another plateau is observed for self-correlation function
$F_s(q,t)$ (Fig.~\ref{fig:fig3})(b) in the time interval
corresponding to the caging regime ($\beta$-relaxation regime in
mode-coupling theory \cite{Gotze,Gotze99,Das}).The correlation
functions start to decay from the plateau at times corresponding
to the onset of the diffusive regime in the MSD
($\alpha$-relaxation regime). In order to estimate the transition
temperature, we calculate diffusivity $D$ as the long time limit
of $<\Delta r(t)^2>/6t$. As  predicted by Mode-Coupling Theory
(MCT), in the vicinity of a glass transition point there is the
power-law temperature dependence of the diffusivity $D\propto
(T-T_c)^{\gamma}$. In accordance with MCT
\cite{Gotze,Gotze99,Das}, $\gamma \geqslant 1.75$. From
Fig.~\ref{fig:fig3}(a) one has $T_c=0.079$ and $\gamma=2.29$.

The apparent glass-transition temperature is above the melting
point of the low-density FCC and FCO phases (see
Fig.~\ref{fig:fig4}(d)). This suggests that  the ``glassy'' phase
that we observe is thermodynamically stable. This is rather
unusual for one-component liquids. In simulations, glassy behavior
is usually observed in metastable mixtures (see
e.g.~\cite{KobAndersen,Kob,Moreno}) where crystal nucleation is
kinetically suppressed. One could argue that, in the glassy
region, the present system behaves like a ``pseudo-binary''
mixture of spheres with diameters $d$ and $\sigma_s$ and that the
freezing-point depression is analogous to that expected in a
binary system with a eutectic point: there are some values of the
diameter ratio such that crystalline structures are strongly
unfavorable and the glassy phase is stable even for very low
temperatures. The glassy behavior in the reentrant liquid
disappears at higher temperatures. In this density range we also
found the anomalous behavior of isochores which corresponds to the
negative thermal expansion coefficient. This behavior will be
discussed in detail in a subsequent publication.

The behavior that we observe in this three-dimensional system
bears some striking similarities to that observed in
two-dimensional system with ``repulsive-shoulder''
potentials~\cite{Camp,MalescioPellicane,Glaser,NorizoeKawakatsu}.
These simulations suggest that such systems have a low-temperature
reentrant liquid phase where particles intend to arrange in more
or less complex, locally ordered patterns. Norizoe and
Kawakatsu~\cite{NorizoeKawakatsu} find evidence for local
clustering and the formation of percolating clusters in the
reentrant liquid phase of a 3D repulsive-shoulder model. Moreover,
refs.~\cite{Camp,MalescioPellicane,Glaser,NorizoeKawakatsu} find
that diffusion in this part of the phase diagram is very slow -
for this reason, they refer to these percolating cluster phases as
``amorphous solids''. However, none of these studies include
free-energy calculations and hence the equilibrium phase
boundaries between solid and reentrant liquid are not known. Our
free energy calculations support that the notion that these
amorphous solids are thermodynamically stable.

\begin{figure}
\includegraphics[width=4cm]{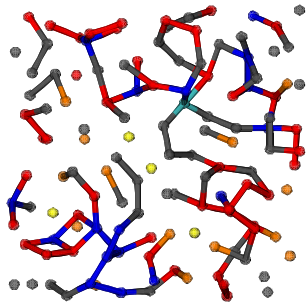}%
\caption{\label{fig:fig6} (Color online). Snapshot of the
disorderded phase for the $\sigma_s=1.35$, for $\rho=0.55$,
$T=0.1$. The colors on the snapshots correspond to the number of
the nearest neighbours (yellow - 0, orange - 1, dark grey - 2, red
- 3, blue - 4, light blue - 5).}
\end{figure}

In the Fig.~\ref{fig:fig6} we represent the snapshot of our system
in this region for $\sigma_s=1.35$, for $\rho=0.55$, $T=0.1$.
Particles in the snapshot are colored according to the number of
the nearest neighbours. The nearest neighbours  to the particle X
were considered to be all particles within the distance of the
first minimum of the radial distribution function. After reentrant
melting the structure corresponds to an assembly of linear
``strings'' of particles with 2 near neighbours. It is possible
that the structure of liquid here is 3D analog of the spatial
configurations found in \cite{MalescioPellicane}, this point
clearly requires further study.

In summary, we have performed the extensive computer simulations
of the phase behavior of  systems described by the soft, purely
repulsive step potential (\ref{2}) in three dimensions.  We find a
surprisingly complex phase behavior. We argue that the evolution
of the phase diagram  may be qualitatively understood by
considering this one-component system as a pseudo-binary mixture
of large and small spheres. Interestingly,  the phase diagram
includes two crystalline FCC domains separated by a sequence of
the structural phase transitions and a reentrant liquid that
becomes amorphous at low temperatures. The phase behavior of
systems with even wider repulsive steps will be discussed in
separate publication.

\begin{acknowledgments}
We thank V. V. Brazhkin for stimulating discussions. The work was
supported in part by the Russian Foundation for Basic Research
(Grants No 05-02-17280 and No 05-02-17621), the Fund of the
President of Russian Federation for Support of Young Scientists
(MK-2905.2007.2) and NWO-RFBR Grant No 047.016.001. The work of the
FOM Institute is part of the research program of FOM and is made
possible by financial support from the Netherlands organization for
Scientific Research (NWO).

\end{acknowledgments}


\end{document}